# Pressure-induced amorphization and polyamorphism in one-dimensional single crystal TiO$_2$ nanomaterials

*Quanjun Li, Bingbing Liu,* * Lin Wang, Dongmei Li, Ran Liu, Bo Zou, Tian Cui, Guangtian Zou*

State Key Laboratory of Superhard Materials, Jilin University, Changchun 130012, China

*Yue Meng, Ho-kwang Mao*

HPCAT, Carnegie Institution of Washington, Building 434E, 9700 South Cass Avenue, Argonne, Illinois 60439, USA

*Zhenxian Liu*

U2A beamline, Carnegie Institution of Washington, Upton, NY 11973, USA

*Jing Liu*

Institute of High Energy Physics, Chinese Academe of Sciences, Beijing 10023, China

*Jixue Li*

State Key Laboratory of Inorganic Synthesis and Preparative Chemistry, Jilin University, Changchun 130012, China

---

* Corresponding author. E-mail: liubb@jlu.edu.cn. Tel/Fax: 86-431-85168256.

**ABSTRACT:** The structural phase transitions of single crystal $TiO_2$-B nanoribbons were investigated in-situ at high-pressure using the synchrotron X-ray diffraction and the Raman scattering. Our results have shown a pressure-induced amorphization (PIA) occurred in $TiO_2$-B nanoribbons upon compression, resulting in a high density amorphous (HDA) form related to the baddeleyite structure. Upon decompression, the HDA form transforms to a low density amorphous (LDA) form while the samples still maintain their pristine nanoribbon shape. HRTEM imaging reveals that the LDA phase has an α-$PbO_2$ structure with short range order. We propose a homogeneous nucleation mechanism to explain the pressure-induced amorphous phase transitions in the $TiO_2$-B nanoribbons. Our study demonstrates for the first time that PIA and polyamorphism occurred in the one-dimensional (1D) $TiO_2$ nanomaterials and provides a new method for preparing 1D amorphous nanomaterials from crystalline nanomaterials.

**TABLE OF CONTENTS GRAPHIC**

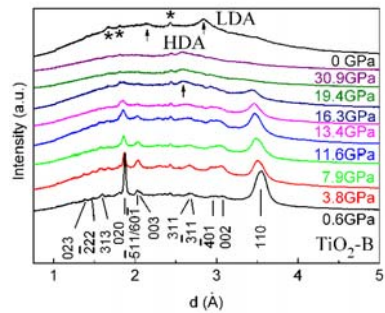

Single crystal $TiO_2$-B nanoribbons transform into high density amorphous (HDA) $TiO_2$ nanoribbons upon compression, in which the one-dimensional confinement plays an important role during the high pressure phase transitions. Upon decompression, the HDA $TiO_2$ nanoribbons converted into low density amorphous (LDA) $TiO_2$ nanoribbons. It provides a new potential method for preparing one-dimensional amorphous nanomaterials.

PIA in the solids has been a subject of intense interest because of its fundamental importance in many fields, including the Earth and planetary sciences, physics, chemistry and material science.[1-5] PIA and the transition from HDA to LDA on decompression have been observed in many classes of materials. The earliest experimental evidence was observed in $H_2O$[6-8], which is related to density-driven phase transitions. Later on, similar phenomena were found in other tetrahedrally coordinated solid Si, Ge, $SiO_2$, and $GeO_2$[2, 9-11]. Recently, it has been found that PIA and the HDA—LDA transition occurs in octahedrally coordinated $TiO_2$ nanoparticles of a particular anatase phase. Lately, several studies have focused on these anatase-phase nanoparticles, investigating the mechanism of PIA and polyamorphism[12-14], especially determining the amorphous structure of the HDA and LDA forms. It has been proposed that the HDA and LDA forms are structurally related to the baddeleyite and α-$PbO_2$ structures and that Ti-O coordination number plays an important role in this phenomenon[13,15]. In addition, the effect of the morphology of nanomaterials upon their high pressure behavior has become a new topic[17-19]. Tetrahedrally coordinated solid Si nanowires have different high pressure behaviors from the bulk and nanoparticle because of their specific geometric morphology[20], however, no PIA was found in these systems. Whether PIA and polyamorphism occur in tetrahedrally and octahedrally coordinated one-dimensional nanostructures is not clear. It has been known that $TiO_2$-B is a metastable $TiO_2$ polymorph, which is a monoclinic structure and composed of corrugated sheets of edge- and corner-sharing $TiO_6$ octahedra. It is octahedrally coordinated but has a different structure from the anatase phase. It is expected to be a new candidate for studying PIA and polyamorphism in octahedrally coordinated nanomaterials. This motivated us to further study the high pressure induced phase transitions for single crystal $TiO_2$-B nanoribbons, exploring PIA and polyamorphism in the 1D nanomaterials.

In this work, we report the first PIA and HDA—LDA transition found in one dimensional $TiO_2$ nanomaterials, which further extend the investigated range for PIA and polyamorphism. The essence of the polyamorphism is revealed directly by HRTEM and it is found that the short-range ordered structure in the LDA form is α-PbO$_2$ in TiO$_2$ nanoribbons. We suggest a homogeneous nucleation mechanism for the high pressure phase transition in the 1D TiO$_2$-B nanoribbons.

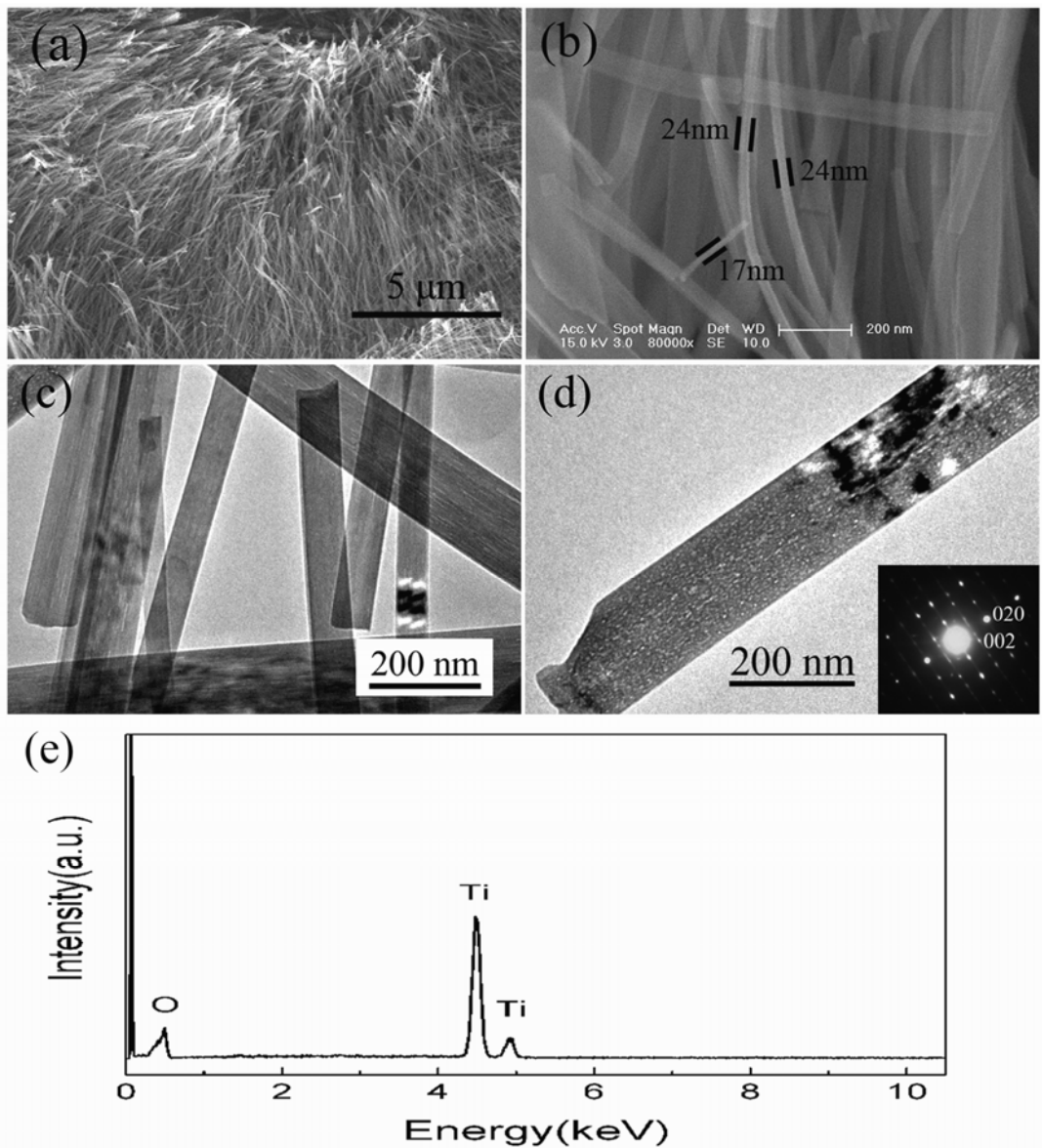

**Figure 1.** Low-magnification (a) and high-magnification (b) FESEM image of the TiO$_2$-B nanoribbons. (c) Low-magnification TEM image of the TiO$_2$-B nanoribbons. (d) High-magnification TEM image of a single TiO$_2$-B nanoribbon together with the corresponding SAED in the inset. (e) EDS spectrum of the TiO$_2$-B nanoribbons.

Figure 1 shows the SEM and TEM images of the TiO$_2$-B nanoribbons. An overview SEM image is presented in Figure 1a, which reveals the presence of a large number of highly uniform nanoribbons with lengths of tens of micrometers. From Figure 1b, it can be seen clearly that the ribbon-like product with widths in the range of 50-200 nm and thickness of ~20 nm is formed. TEM images (Figure 1c and d) further confirm the ribbon-like morphology. They are in good accordance with Figure 1b. An individual nanoribbon is shown in Figure 1d, the selected-area electron diffraction (SAED) pattern (inset of Figure 1d) reveals that the nanoribbon is a single crystal of layered $TiO_2$-B grown along the [010] direction. We further investigated the composition of the formed nanoribbons using EDS. Strong Ti and O peaks are seen in Figure 1e, indicating the formed nanoribbons are pure $TiO_2$.

The pressure evolution of the energy dispersive X-ray diffraction patterns of $TiO_2$-B nanoribbons are shown in Figure 2. The diffraction peaks are all indexed to the $TiO_2$-B phase, in which the peak of the (020) plane is obviously sharp. It also indicates that the $TiO_2$-B nanoribbons are preferentially grown along the [010] direction. With increasing pressure, all the peaks of $TiO_2$-B shift to smaller d-spacing, and as the pressure exceeds 13.4 GPa, the broadening of the diffraction peaks became very obvious. At pressure > 16.3 GPa, all of diffraction peaks disappeared. Except for that, a relatively broad and weak new band appeared and it remained up to the highest pressure of 31 GPa. These results indicate that an amorphization in the $TiO_2$-B nanoribbons has occurred in the pressure range of 16.3-19.4 GPa. It also demonstrates that the sample underwent a structural transformation, however, not to a high-pressure crystalline phase as observed in the bulk sample.[21] Instead, an amorphous phase was observed, which shows an amorphous diffraction peak at ~2.58 Å corresponding to the (111) plane of the baddeleyite structure. We further studied the sample released from the highest pressure and compared the sample to the diffraction patterns for the sample at 30.9 GPa. Strikingly two broad peaks were observed at ambient pressure (Figure 2b), which is distinctly different from the high pressure amorphous phase at 30.9 GPa. This indicates a low pressure amorphous phase is formed upon decompression. The center of the two broadened peaks are at 2.13 and 2.84 Å in low pressure amorphous sample, respectively, which corresponding to the (121) and (111) plane of α-$PbO_2$ phase. These data suggest that the high pressure amorphous phase and the low pressure amorphous phase have a relationship to α-$PbO_2$ and baddeleyite, respectively. According to the previous studies of $TiO_2$, $H_2O$, and Si,[2,5,6,9,12,13,15] it is reasonable to conclude that the transition of the high pressure amorphous phase to the low pressure amorphous phase is also the transition of an HDA form to an LDA form. It is to be noted that polyamorphism also exists in $TiO_2$-B nanoribbons in addition to anatase nanoparticles.

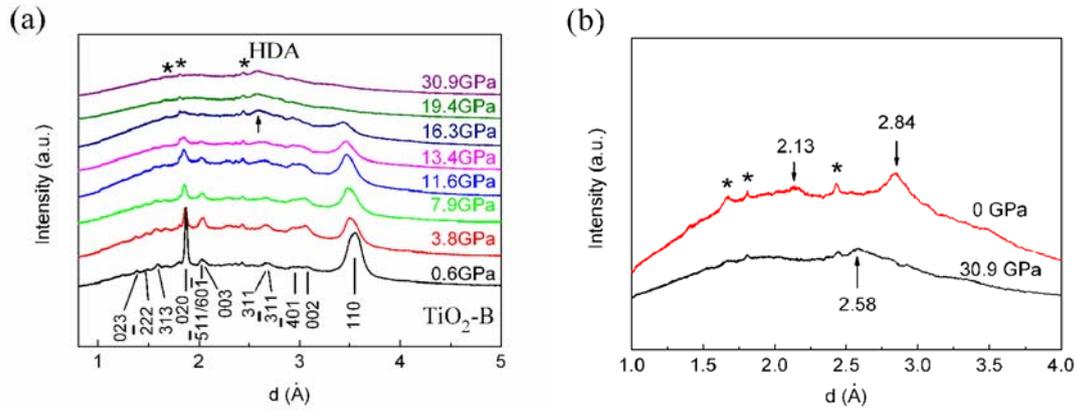

**Figure 2.** (a) High pressure powder X-ray diffraction patterns of $TiO_2$-B nanoribbons up to 30.9 GPa at room temperature. (b) A comparison between the x-ray patterns of the as-synthesized $TiO_2$-B nanoribbons obtained at 30.9 GPa and recovered at ambient conditions, respectively. (Three weak peaks (marked with asterisk) are derived from the energy-dispersive synchrotron x-ray diffraction system).

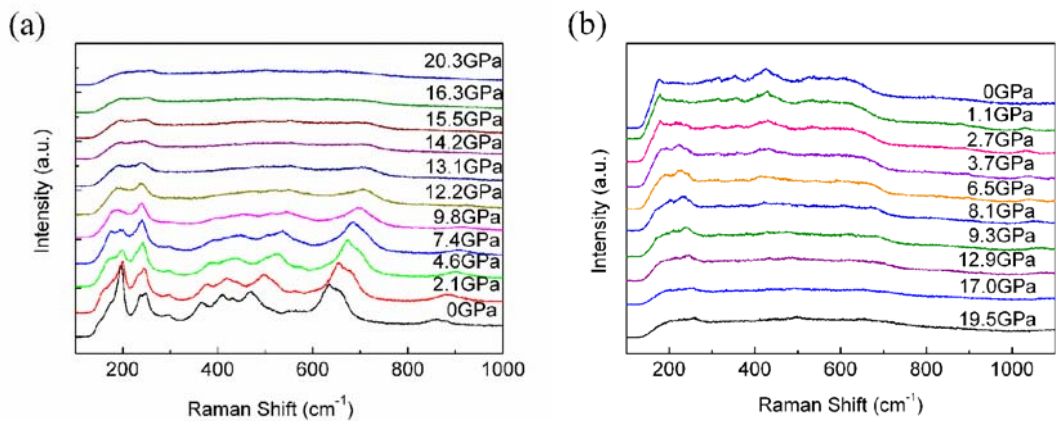

**Figure 3.** (a) Raman spectra of $TiO_2$-B nanoribbons measured during compression and decompression. (a) Compression. (b) Decompression.

To further verify the pressure-induced amorphization and polyamorphism for the $TiO_2$-B nanoribbons, we also carried out high-pressure Raman measurements. Figure 3 shows the Raman spectra of the $TiO_2$-B nanoribbons measured during compression and decompression. As shown in Figure 3a, with increasing pressure to about ~16 GPa, all Raman modes of $TiO_2$-B disappear. In stead, two broad weak bands near 220-280 cm$^{-1}$ and 475-530 cm$^{-1}$ start to arise and remain up to 20.3 GPa which are similar to the Raman spectra of HDA TiO$_2$[13]. It means that PIA occurs at pressure of about ~16 GPa and forms a HDA phase. During decompression, the intensity of the two bands (220-280 cm$^{-1}$ and 475-530 cm$^{-1}$) enhance with decreasing pressure (Figure 3b). When the pressure released to ~8.1 GPa, new broad features began to emerge at bands of 180-210 cm$^{-1}$ and 350-440 cm$^{-1}$. Other new broad weak bands of 300-320 cm$^{-1}$ and 340-370 cm$^{-1}$ occur at ~2.7 GPa. At the same time, the Raman spectra of HDA phase disappear completely. These new broad peaks (180-210 cm$^{-1}$, 300-320 cm$^{-1}$, 340-370 cm$^{-1}$, and 350-440 cm$^{-1}$) grow and redshift with decreasing pressure and remain after decompression to ambient pressure. These results further verify that the HDA phase transforms into the LDA phase during decompression.

It is known that bulk TiO$_2$-B transform into anatase structure at 6 Gpa.[21] Obviously, this high pressure phase transition of TiO$_2$-B nanoribbons is quite different from that of the corresponding bulk material, not only is the critical transiton pressure higher than that of the bulk sample, enhancing the stability, but also high pressure induced amorphization occurs. We suggest that the enhanced stability for TiO$_2$-B nanoribbons in our case is due to the fact that the nanoribbons are regular extended solids along the length and width of the structure, but have nanometer-scale thicknesses (~20 nm). The structure of TiO$_2$-B is composed of corrugated sheets of edge- and corner-sharing TiO$_6$ octahedra,[20] which contain chains of edge-sharing octahedra in one orientation. TiO$_2$-B thus structurally relates to anatase, which is different from the other polymorphs having chains of edge-sharing octahedra in two orientations. As proposed in the previous literatures for anatase nanoparticles,[12,13,22] the critical size of nucleation and growth for high pressure phases determines the PIA and polyamorphization in nanoparticles. Considering the difference in morphology between our sample and that of nanoparticles, we consider that the one-dimensional confinement in thickness for the nanoribbons precludes the nucleation and growth of other crystalline phases (anatase or the other high pressure phases of TiO$_2$) under high pressure which are not energetically favorable.

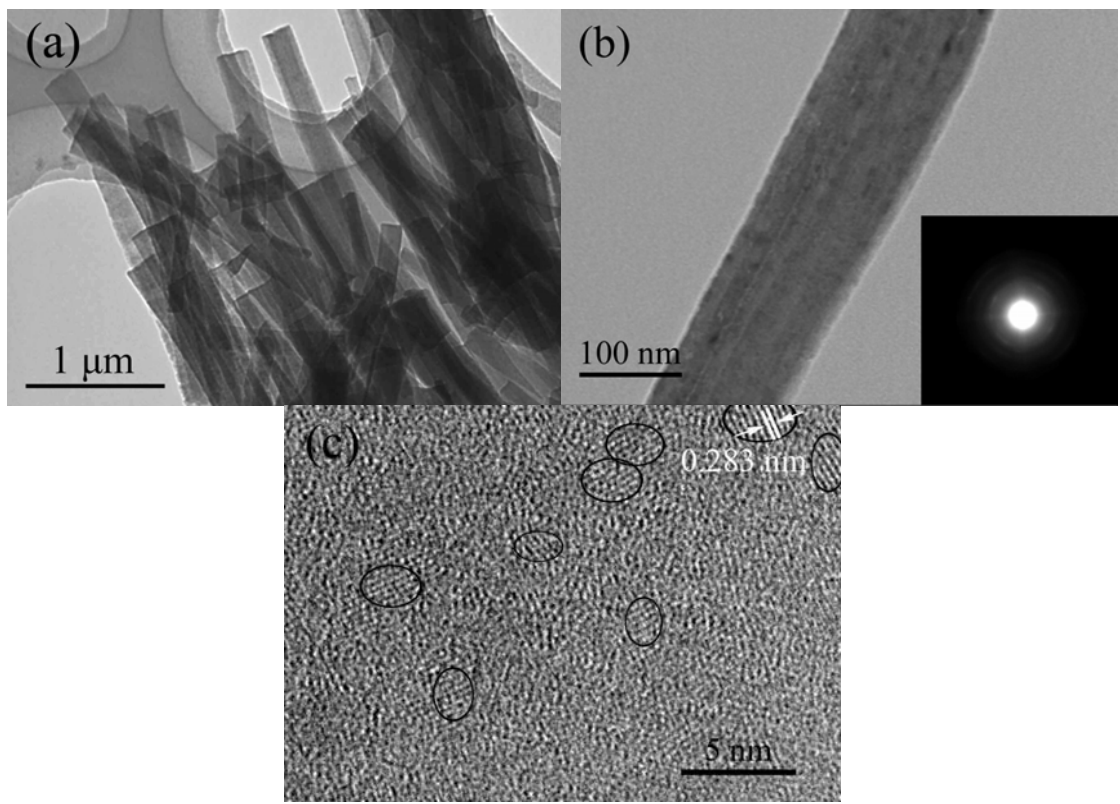

**Figure 4.** TEM image of the nanoribbons after released from 31 GPa to ambient pressure. (a) typical TEM image of LDA TiO$_2$ nanoribbons, (b) an individual LDA TiO$_2$ nanoribbons and its SAED (inset image), (c) HRTEM image of an individual LDA TiO$_2$ nanoribbon.

To further reveal the mechanism of the PIA and polyamorphism and to observe the morphology change of the nanoribbons, as well as to give a direct picture of the microstructure of the LDA form, we have carried out TEM and HRTEM analysis of the recovered sample quenched from high pressure through a delicate experimental operation. As we know, it is very difficult to observe the structures and the morphologies of samples recovered from a diamond anvil cell because the sample is rather subtle for electron microscopy analysis, so there are only a few TEM/HRTEM investigations in previous in situ high pressure studies.[16] Here, the TEM and HRTEM images of the sample are shown in Figure 4. It is clear that the sample (LDA form) still maintains its ribbon-like morphology without obvious shape change after the phase transitions (TiO$_2$-B—HDA—LDA) (Figure 4a). The SAED pattern of the LDA polyamorph is shown in the inset of Figure 4b. Clear and distinct diffraction rings are not observed, suggesting that the nanoribbons are amorphous. To further probe the disorder nature of the LDA nanoribbons, the HRTEM image of an individual nanoribbon is given in Figure 4c. Obviously, long-range ordered structure does not exist in the LDA nanoribbons, but some short-range ordered domains with diameters about 1-3 nm are distributed uniformly inside the nanoribbons. These domains have no identical growth orientation. The lattice fringes of the domains show a distance of about 0.283 nm, which correspond to the (111) plane of the α-$PbO_2$ phase. It is in good agreement with our XRD and Raman results (Figure 2b and Figure 3b) that LDA form was obtained upon decompression. This image shows a representative picture of the amorphous structure which is strikingly different from the LDA of $TiO_2$ nanoparticles induced by PIA in anatase $TiO_2$ nanoparticles. In anatase nanoparticles, the high pressure crystalline phase does not form because their three-dimensional finite size is estimated to be less than the critical size (~12 nm) for nucleation and growth of the high pressure phase.[22] Compared with this, in our case we can clearly see the nucleation in single-crystalline $TiO_2$-B nanoribbons, the short-range ordered domains with a couple of nanometers are formed which can be considered as the nucleus for the high pressure phase of α-$PbO_2$. This image provides direct evidence of an amorphous structure with α-$PbO_2$ phase as the short-range domains. In addition, it is worth noting that the morphology of the nanoribbons remains during the high pressure phase transition. This is different from the changes of morphology observed in other 1D materials quenched from high pressure, such as ZnS nanobelts and CdSe nanorods,[16,17] in which the 1D samples are fractured after undergoing crystalline phase transformation. This result indicates that 1D amorphous $TiO_2$ nanomaterials can be obtained by high pressure treatment and the high surface energy of the nanoribbons likely prohibits the change of morphology during the phase transitions.

According to the HRTEM results, here we propose a homogeneous nucleation mechanism for the high pressure phase transition in the single crystal $TiO_2$-B nanoribbons and tentatively explain the structural relation between the LDA/HDA and α-$PbO_2$/baddeleyite forms. As depicted in Figure 3c, these short-range ordered domains or "nucleus" with an α-$PbO_2$ form give direct evidence for the homogeneous nucleation. We thus deduce that the α-$PbO_2$ nucleus come from the high pressure phase of the baddeleyite nucleus upon decompression. In the single crystal $TiO_2$-B nanoribbons, the baddeleyite nucleus occurs uniformly in the body of the nanoribbons when the pressure is higher than 16 GPa, however, these nucleation sites are difficult to further grow into large grains due to the confinement effect in the thickness of the nanoribbons, which results in the formation of HDA forms with a structural relation to baddeleyite. Upon decompression, these baddeleyite nucleation sites inside the HDA form transform to the α-$PbO_2$ nucleus, which induces the phase transition of HDA-LDA. The baddeleyite and α-$PbO_2$ structures of $TiO_2$ are 7 and 6 Ti-O coordination numbers, respectively.[13] The baddeleyite—α-$PbO_2$ phase transition involves a coordination number change from 7 to 6, which has been used to explain the density differences of HDA/LDA polyamorphs that are produced via PIA from anatase nanoparticles.[13] It is reasonable to presume that a similar situation occurs in our case. The pressure induced HDA-LDA transition is related to the nucleus of the high-pressure crystalline phase. That is, the baddeleyite nucleus and the α-$PbO_2$ nucleus exist in the HDA and LDA nanoribbons, respectively, which dominate the pressure induced HDA-LDA transition between the two polyamorphic forms.

In summary, the high-pressure induced structural phase transition of 1D $TiO_2$-B nanoribbons has been investigated at high-pressure by the synchrotron radiation X-ray diffraction and the Raman scattering. It was found that PIA and the amorphous phase transition from HDA to LDA occur in $TiO_2$-B nanoribbons, which are attributed to their particular 1D ribbon-like morphology. We revealed the LDA phase quenched from high pressure is a long-range disordered and short-range ordered α-$PbO_2$ structure by a direct HRTEM observation. The morphologies of the nanoribbons were retained during the phase transitions of $TiO_2$-B—HDA—LDA. We propose a homogeneous nucleation mechanism to explain the high pressure induced phase transitions for the $TiO_2$-B nanoribbons. Our study provides a first example of PIA and polyamorphism occurring in 1D $TiO_2$ nanomaterials and also provides a new method for preparing 1D amorphous nanomaterials from crystalline nanomaterials, which extends polyamorphic systems for studying HDA—LDA transformations.

**Experimental Section**

Single crystalline $TiO_2$-B nanoribbons were synthesized by a simple hydrothermal reaction between NaOH and $TiO_2$ nanoparticles (Degussa P25) as described in our previous work[23], but at a higher reaction temperature (200 °C). The sample was characterized using FESEM (15 KV, XL 30, ESEM), transmission electron microscopy (TEM) (200 KV, HITACHI, H-8100IV), and high resolution transmission electron microscopy (HRTEM) (JEOL JEM-3010). In situ energy-dispersive synchrotron X-ray diffraction measurements under high pressure were carried out at the Advanced Photon Source (APS). Part of the X-ray diffraction measurement has been repeated at the 4W2 High-Pressure Station of Beijing Synchrotron Radiation Facility (BSRF). In addition, high pressure Raman measurements were carried out at room temperature by a Renishaw inVia Raman spectrometer with a He-Ne 633 nm laser. High pressure was generated by a diamond anvil cell (DAC) with 4:1 methanol-ethanol mixture as the pressure-transmitting medium. The pressures were determined from the pressure dependent shift of the R1 line fluorescence of ruby. After released to ambient pressure, took out the top of DAC and dropped a little ethanol into the hole of gasket, sucked the sample (LDA $TiO_2$ nanoribbons) out the hole by a tiny suction pipe, then dispersed on a carbon-coated copper grid for TEM and HRTEM observation.

**ACKNOWLEDGMENT.** This work was supported financially by NSFC (10674053, 20773043, 10574053), the National Basic Research Program of China (2005CB724400, 2001CB711201), the Program for Changjiang Scholar and Innovative Research Team in University (IRT0625), the 2007 Cheung Kong Scholars Programme of China, and the National Found for Fostering Talents of basic Science (J0730311). The use of the National Synchrotron Light Source beamlines X17C and U2A are supported by NSF COMPRES EAR01-35554 and by US-DOE contract DE-AC02-10886.

**References**

(1) Mishima, O.; Calvert, L. D.; Whalley, E. 'Melting ice' I at 77K and 10 kbar: a New Method of Making Amorphous Solids. *Nature*, **1984**, *310*, 393-395.

(2) Deb, S. K.; Wilding, M.; Somayazulu, M.; McMillan, P. F. Pressure-Induced Amorphization and an Amorphous-Amorphous Transition in Densified Porous Silicon. *Nature*, **2001**, *414*, 528-530.

(3) Perottoni, C. A.; Da Jornada, J. A. H. Pressure-Induced Amorphization and Negative Thermal Expansion in $ZrW_2O_8$. *Science*, **1998**, *280*, 886-889.

(4) Hemley, R. J.; Jephcoat, A. P. ; Mao, H. K.; Ming, L. C.; Manghnan, M. H. Pressure-Induced Amorphization of Crystalline Silica. *Nature*, **1988**, *334*, 52-54.

(5) McMillan, P. F. Polyamorphic Transformations in Liquids and Glasses. *J. Mater. Chem.* **2004**, *14*, 1506-1512.

(6) Mishima, O.; Calvert, L. D.; Whalley, E. An Apparently First-Order Transition between Two Amorphous Phases of Ice Induced by Pressure. *Nature*, **1985**, *314*, 76-78.

(7) Jenniskens, P.; Blake, D. F. Structural Transitions in Amorphous Water Ice and Astrophysical Implications. *Science*, **1994**, *256*, 753-756.

(8) Tulk, C. A.; Benmore, C. J.; Urquidi, J.; Klug, D. D.; Neuefeind, J.; Tomberli, B.; Egelstaff, P. A. Structural Studies of Several Distinct Metastable Forms of Amorphous Ice. *Science*, **2002**, *297*, 1320-1323.

(9) McMillan, P. F.; Wilson, M. Daisenberger, D.; Machon, D. A Density-Driven Phase Transition between Semiconducting and Metallic Polyamorphs of Silicon. *Nat. mater.* **2005**, *4*, 680-684.

(10) Di Cicco, A.; Congeduti, A.; Coppari, F.; Chervin, J. C.; Baudelet, F.; Polian, A. Interplay between Morphology and Metallization in Amorphous-Amouphous Transitions. *Phys. Rev. B* **2008**, *78*, 033309.

(11) Itie, J. P.; Polian, A.; Calas, G.; Petiau, J.; Fontaine, A.; Tolentino, H. Pressure-Induced Coordination Change in Crystalline and Vitreous $GeO_2$. *Phys. Rev. Lett.* **1989**, *63*, 398-401.

(12) Swamy, V.; Kuznetsov, A.; Dubrovinsky, L. S.; Caruso, R. A.; Shchukin, D. G.; Muddle, B. C. Finite-Size and Pressure Effects on the Raman Spectrum of Nanocrystalline Anatase $TiO_2$. *Phys. Rev. B* **2005**, *71*, 184302.

(13) Swamy, V.; Kuznetsov, A.; Dubrovinsky, L. S.; McMillan, P. F.; Prakapenka, V. B.; Shen, G.; Muddle, B. C. Size-Dependent Pressure-Induced Amorphization in Nanoanatase and Eventual Pressure-Induced Disorder on the Nanometer Scale. *Phys. Rev. Lett.* **2006**, *96*, 135702.

(14) Pischedda, V.; Hearne, G. R.; Dawe, A. M.; Lowther, J. E. Ultrastability and Enhanced Stiffness of ~6 nm $TiO_2$ Nanoanatase and Eventual Pressure-Induced Disorder on the Nanometer Scale. *Phys. Rev. Lett.* **2006**, *96*, 035509.

(15) Flank, A. M.; Lagarde, P.; Itie, J. P.; Polian, A.; Hearne, G. R. Pressure-Induced Amorphization and a Possible Polyamorphism. *Phys. Rev. B* **2008**, *77*, 224112.

(16) Marchand, R.; Brohan, L.; Tournoux, M. $TiO_2$(B) a New Form of Titanium Dioxide and the Potassium Octatitanate $K_2Ti_8O_{17}$. *Mat. Res. Bull.* **1980**, *15*, 1129-1133.

(17) Wang, Z. W.; Daemen, L. L.; Zhao, Y. S.; Zha, C. S.; Downs, R. T.; Wang, X. D.; Wang, Z. L.; Hemley, R. Morphology-Tuned Wurtzite-Type ZnS Nanobelts. J. *Nat. Mater.* **2005**, *13*, 1-6.

(18) Zaziski, D.; Prilliman, S.; Scher, E.; Casula, M.; Wickham, J.; Clark, S. M.; Alivisatos, A. P. Critical Size for Fracture during Solid-Solid Phase Transformations. *Nano. Lett.* **2004**, *4*, 943-946.

(19) Shen, L. H.; Li, X. F.; Ma, Y. M.; Yang, K. F.; Lei, W. W.; Cui, Q. L.; Zou, G. T. Pressure-Induced Structural Transition in AlN Nanowires. *Appl. Phys. Lett.* **2006**, *89*, 141903.

(20) Wang, Y. J.; Zhang, J. Z.; Wu, J.; Coffer, J. L.; Lin, Z. J.; Sinogeikin, S. V.; Yang, W. G.; Zhao, Y. S. Phase Transition and Compressibility in Silicon Nanowires. *Nano. Lett.* **2008**, *8*, 2891-2895.

(21) Brohan, L.; Verbaere, A.; Tournoux, M.; Demazeau, G. La Transformation $TiO_2$(B)-Anatase. *Mat. Res. Bull.* **1982**, *17*, 355-361.

(22) Hearne, G. R.; Zhao, J.; Dawe, A. M.; Pischedda, V.; Maaza, M.; Nieuwoudt, M. K.; Kibasomba, P.; Nemraoui, O.; Comins, J. D.; Witcomb, M. J. Effect of Grain Size on Structural Transitions in Anatase $TiO_2$: A Raman Spectroscopy Study at High Pressure. *Phys. Rev. B* **2004**, *70*, 134102.

(23) Li, Q. J.; Zhang, J. W.; Liu, B. B.; Li, M.; Yu, S. D.; Wang, L.; Li, Z. P.; Liu, D. D.; Hou, Y. Y.; Zou, Y. G.; Zou, B.; Cui, T.; Zou, G. T. Synthesis and Electrochemical Properties of TiO$_2$-B@C Core-Shell Nanoribbons. *Cryst. Growth Des.* **2008**, *8*, 1812-1814.